\documentclass[aps,pra,10pt,twocolumn,groupedaddress,showpacs,preprintnumbers,amsmath,amssymb, longbibliography]{revtex4-2}

\usepackage{graphicx}
\usepackage{color}
\usepackage{dcolumn}
\usepackage{times}
\usepackage{bm}
\usepackage{verbatim}
\usepackage{mathbbol}
\usepackage{amsmath}
\usepackage{amssymb}
\usepackage{yfonts}
\usepackage{physics}
\usepackage{mathrsfs}
\usepackage{upgreek}
\usepackage{soul}
\usepackage{xcolor}
\setstcolor{red}
\usepackage[normalem]{ulem} 
\usepackage[caption=false]{subfig}
\usepackage{float}
\definecolor{dark_red}{rgb}{0.75,0,0}
\definecolor{dark_purple}{rgb}{0.75,0,0.75}
\definecolor{dark_blue}{rgb}{0,0,0.75}
\definecolor{dark_green}{rgb}{0,0.60,0}
\usepackage[colorlinks,citecolor=blue,linkcolor=blue]{hyperref}
\usepackage{contour}

\def\XXint#1#2#3{{\setbox0=\hbox{$#1{#2#3}{\int}$}
     \vcenter{\hbox{$#2#3$}}\kern-.5\wd0}}

\begin{document}

\title{Non-perturbative Effects in Attosecond Four-Wave Mixing Spectra}

\author{Sergio Yanez-Pagans}
\affiliation{Department of Physics, University of Arizona, Tucson, Arizona 85721, USA}
\author{Nathan Harkema}
\affiliation{Department of Physics, University of Arizona, Tucson, Arizona 85721, USA}
\author{Arvinder Sandhu}
\email{asandhu@arizona.edu}
\affiliation{Department of Physics, University of Arizona, Tucson, Arizona 85721, USA}

\author{Coleman Cariker}
\affiliation{Department of Physics, University of Central Florida, Orlando, Florida 32816, USA}
\author{Luca Argenti}
\email{luca.argenti@ucf.edu}
\affiliation{Department of Physics, University of Central Florida, Orlando, Florida 32816, USA}

\date{\today}


\begin{abstract}
We study the nonlinear optical response of argon to a four-wave-mixing pulse sequence consisting of an extreme ultraviolet pulse, an overlapping collinear IR and an non-collinear delayed IR pulses. Absorption of an extreme ultraviolet and an IR photon from the collinear beams excites, sequentially, the $3s^{-1}4p$ bright state and the {$3s^{-1}3d/5s$} dark states. The subsequent absorption of an IR photon from the non-collinear beam results in an angled extreme ultraviolet emission whose variation with delay encodes coupling between autoionizing-states, dark-state lifetimes, and non-perturbative effects. 
Both our measurements and \emph{ab initio} simulations of the angled four-wave-mixing signal show a double-peak structure in delay dependence, in excellent agreement with each other. We attribute the minimum at the center of the signal to the rapid Rabi cycling, driven by the IR pulse, between dark states and the $3s^{-1}4p$ resonance, which results in the destructive interference in the final transition amplitude.

\end{abstract}

\maketitle

\section{Introduction}
Attosecond transient absorption spectroscopy (ATAS), which monitors the spectral change of an extreme ultraviolet pulse (XUV) as a function of the delay of an IR pulse, has emerged as a useful tool for probing and controlling ultrafast electronic dynamics in atoms, molecules, and continuous media \cite{goulielmakis2010real,gruson2016attosecond,jimenez2014modulation,argenti2015dressing,kotur2016spectral,chu2013absorption,chu2012photoabsorption,beck2015probing,bernhardt2014high,stooss2019xuv,chew2018attosecond}. ATAS reveals the excitation, radiative coupling, and time-resolved evolution of the target bright excited states. Autoionizing states, in particular, exhibit an asymmetric absorption profile, due to the interference between direct-ionization and excitation to ametastable state followed by Auger decay~\cite{fano1965interpretation}. In ATAS, a probe IR dressing field can be used to control the phase of the metastable states, thus altering their spectral lineshapes and lifetimes~\cite{OttScience2013, ArgentiPRA2015,beck2015probing, beck2014attosecond}, as well as to reconstruct the relative phase of multiple coherently excited metastable states~\cite{OttNature2014}.

Non-Collinear Four-Wave Mixing (NCFWM), an extension of transient absorption spectroscopy, employs two independent IR pulses in a non-collinear alignment to spatially resolve signals emerging from different transition pathways~\cite{cao2016noncollinear}. NCFWM, therefore, allows one to study transition amplitudes from bright to dark metastable states free from the background one-photon absorption signal~\cite{bengtsson2017space}. When operated in perturbative regime, this all-optical technique is capable of accurately measuring the lifetime of intermediate dark autoionizing resonances. Recently, NCFWM was used to measure the decay of dark bound states in argon~\cite{cao2016noncollinear}, and of dark autoionizing states in neon~\cite{PuskarPRA2023}. This technique, however, is not limited to the perturbative regime and can be used to interrogate dark autoionizing states even when they are strongly coupled to neighboring bright resonances.

Our recent ATAS investigation of argon in the proximity of the $3s^{-1}$ threshold, in particular, has revealed a rich non-perturbative dynamics between the autoionizing states of this atom~\cite{HarkemaPRL2021,Yanez-PagansPRA2022}.
In this work, we study how this dynamics is reflected in the NCFWM spectrum monitoring the amplitude of the $3s^{-1}5s$ and $3s^{-1}3d$ dark autoionizing states. 
\begin{figure}[hbtp!]
\includegraphics[width=0.8\columnwidth]{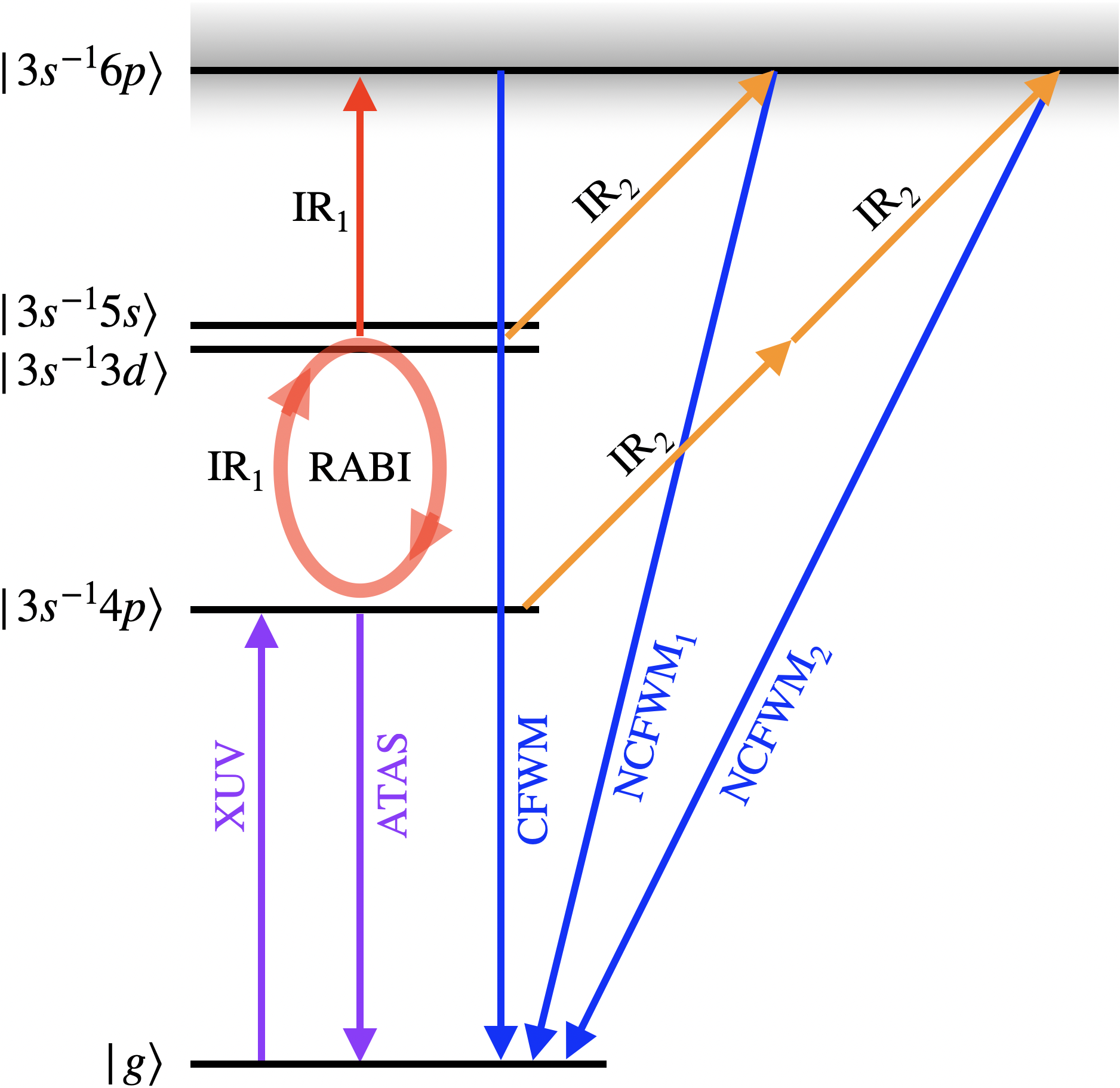}
\caption{\label{fig:EnergyLevels} Scheme of the levels studied in this work. The XUV pump pulse excites the bright $3s^{-1}4p$ level, which the collinear dressing laser $\textsc{ir}_1$ couples to the nearly degenerate $3s^{-1}5s$ and $3s^{-1}3d$ dark states. The forward dipole emission from the laser-dressed $3s^{-1}4p$ state results in the ATAS signal. The angled probe pulse $\textsc{ir}_2$ excites the $3s^{-1}6p$ state, which radiates off axis, giving rise to the NCFWM signals.}
\end{figure}
To interpret the experimental results, we simulate the experiment with the newstock \emph{ab~initio} atomic ionization program~\cite{HarkemaPRL2021,Yanez-PagansPRA2022}. From the single-atom response of the atom to the pump-probe sequence with arbitrary delays for the angled photon, we determine the spatially-resolved collective response of the whole interaction region.
These theoretical simulations confirm that the NCFWM signal at large time delays exhibits the characteristic exponential decay from which the lifetimes of argon's transient dark states can in principle be extracted. 
Here, however, we focus on the regime where the intense dressing pulse and the angled probe IR pulse overlap. In this case, the angled laser probes the non-perturbative dynamics of the autoionizing states under the influence of the dressing laser, as illustrated in Fig.~\ref{fig:EnergyLevels}. Both experiment and theory show that, in these conditions, the NCFWM signal bears the signature of the Rabi cycling between dark and bright autoionizing states.

The most striking effect of the strong radiative coupling between autoionizing states is the transient suppression of the NCFWM signal observed when the pump and probe pulses fully overlap, resulting in two peaks whose separation widens as the intensity of the dressing laser is increased. In the simulations, the minimum of the NCWF signal occurs when the angled laser probes the $3s^{-1}5s/3d$ dark states while their amplitudes change sign, on account of their Rabi oscillation with the bright $3s^{-1}4p$ state. Indeed, the NCFWM signal is most pronounced at the onset and the wind down of the Rabi oscillations, when the angled probe pulse is at the rising and decreasing edges of the dressing pulse, corresponding to two distinct peaks. The~\emph{ab initio} simulations reproduce well both the signal suppression and widening of the time-delay gap between the two peaks as the intensity of the laser is increased, giving credence to our interpretation.

The paper is organized as follows. Sec.~\ref{sec:Experiment} describes the NCFWM experimental setup. Sec.~\ref{sec:Theory} offers a brief summary of the~\emph{ab initio} method used to compute the single-atom response to a sequence of pump-probe pulses. Sec.~\ref{sec:Results} compares the experimental and theoretical findings. In Sec.~\ref{sec:Conclusions} we offer our conclusions. Finally, App.~\ref{app:NCFWMFormula} details how the NCFWM collective signal of the atoms in the interaction region is computed from the optical response of a single-atom.

\section{EXPERIMENTAL SETUP\label{sec:Experiment}}
\begin{figure}[hbtp!]
\includegraphics[width=\columnwidth]{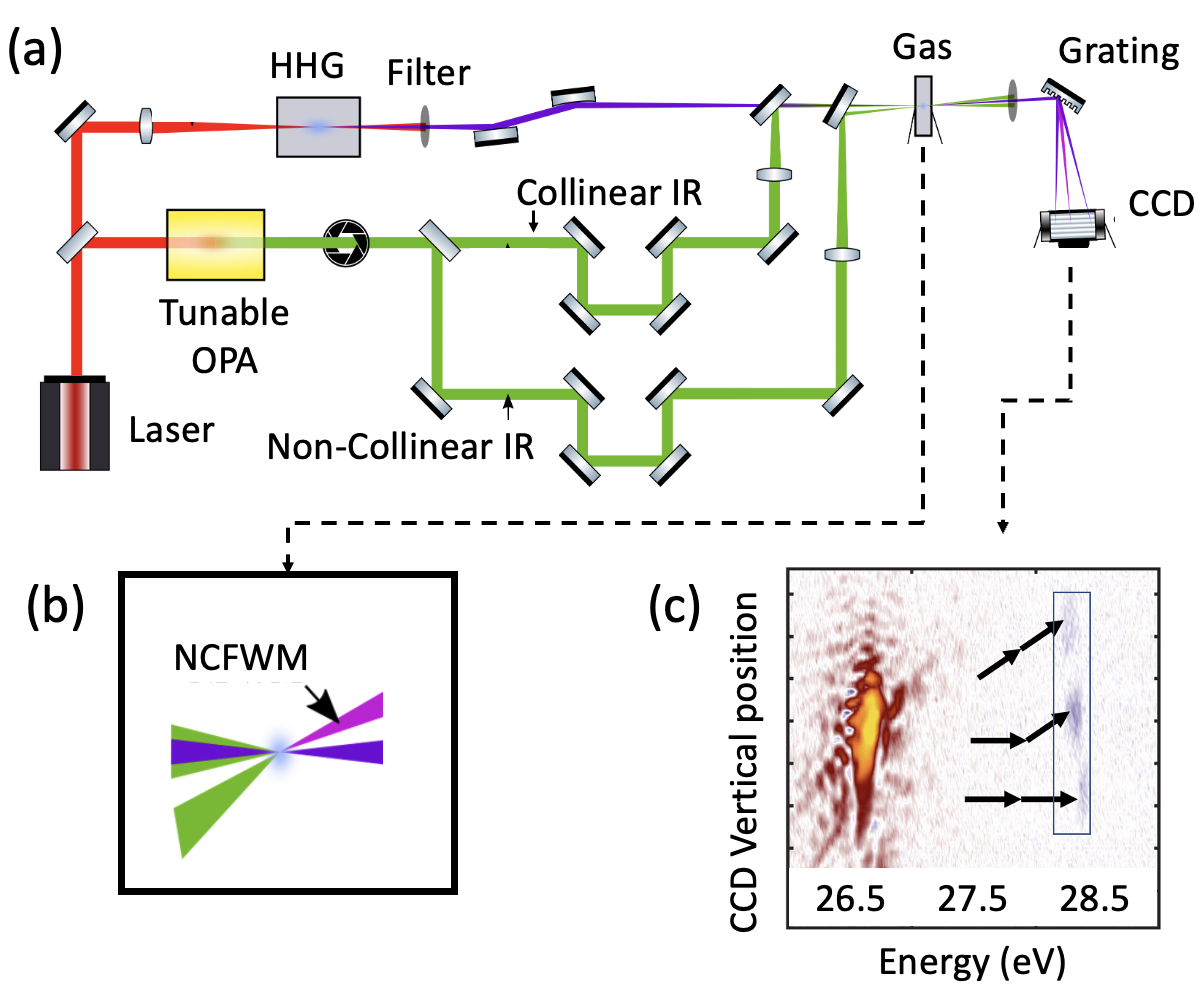}
\caption{\label{fig:ExperimentalSetup} (a) Experimental setup employing HHG for XUV pulse generation, together with a collinear and a non-collinear IR beams, one-collinear and other non-collinear, resultint in the generation of a four-wave-mixing signal in the target gas. (b) The non-collinear IR photon absorption results in XUV emission at an angle relative to the incident beam. (c) Image from the charge-coupled detector (CCD camera)} showing the main XUV harmonic and three FWM signals, each correspong to a different combination of collinear and non-collinear IR photon absorption.
\end{figure}

Our investigations use a three-pulse XUV-IR pump-probe scheme, as shown in Fig.~\ref{fig:ExperimentalSetup}(a). Initially, $\sim$2-mJ 40-fs near-infrared (NIR) pulses are generated by a Ti:Sa laser amplifier at a 1-kHz repetition rate. The 780-nm output beam is divided into two equal parts with a 50/50 beam splitter. The first beam is focused into a xenon-filled gas cell to generate an XUV attosecond-pulse train (APT). The XUV phase-matching conditions are set to maximize the photon flux around 26.62~eV. The second part of the pulsed beam is routed through an optical parametric amplifier (OPA) where the NIR pulse is converted into an infrared (IR) pulse with frequency tunable in the 0.7--1.0~eV energy range. Next, the tunable IR beam is further divided into two parts by means of a second beam splitter. Both arms are routed through delay lines in order to control the relative timing between the XUV and the two IR pulses. 
In addition, half-wave plates and a polarizers are used to separataely control the intensity of the two IR beams. 
A shutter is used to alternate between the XUV only and XUV${+}$IR cases. Subsequently, the XUV and one of the IR pulses are collinearly combined by an annular mirror and focused into a 3-mm-thick argon gas cell (${\sim}$4 Torr). With the XUV and one collinear IR pulse parked at delay zero ($\tau_1{\sim}0$), the second IR pulse with variable delay $\tau_2$ is focused on the reaction region, in a non-collinear geometry. All pulses have linear electric field polarization in the same direction. Finally, the IR beams are filtered out with an aluminum filter and the XUV APT carries on to the XUV spectrometer.

Fig.~\ref{fig:ExperimentalSetup}(b) shows the NCFWM geometry, where the absorption of non-collinear IR photons during XUV interaction, results in new XUV emission which is angled with respect to the incoming XUV beam. The FWM signal of interest in this work, associated to the net absorption of one collinear and one non-collinear IR photon, is emitted in the direction $\vec{k}_{\textsc{ncfwm}}=\vec{k}_{\textsc{xuv}}+\vec{k}_{\textsc{ir}_1}+\vec{k}_{\textsc{ir}_2}$, determined by the phase-matching condition. The XUV emission from the sample is measured using a grating spectrometer and a two-dimensional detector, where we can observe the spectral dispersion along the horizontal axis and emission angle of XUV along the vertical axis (Fig.~\ref{fig:ExperimentalSetup}(c)).  
Apart from the transient absorption change at the main XUV harmonic at energy $~$26.5$~eV$, we see three FWM emission peaks at energy $~$28.5$~eV$, which are displaced along the vertical axis by different amounts. The lowest emission peak along the vertical corresponds to the absorption of XUV alongside two collinear IR photons of $~$1$~eV$ energy each, the middle peak corresponds to the absorption of XUV in conjunction with one collinear and one non-collinear (angled) IR photon, and the peak with largest vertical displacement/angle represents the absorption of XUV plus two non-collinear IR photons. 
We spatially integrate the signal in these three regions and monitor their evolution as a function of the time-delay $\tau_2$ of the angled beam.

\section{Theoretical Method\label{sec:Theory}}
\emph{Ab initio} calculations for the single-atom response of argon are carried out using the NewStock suite of codes~\cite{CarettePRA2013,HarkemaPRL2021,Yanez-PagansPRA2022,CarikerPRA2024}. The localized orbitals are calculated with the MCHF method, using the ATSP2K atomic structure code~\cite{Froese-Fischer1997}, optimizing the active orbitals on the $3p^{-1}$ and $3s^{-1}$ ionic configurations. The configuration space for the neutral argon atom is constructed by augmenting a finite number of MCHF ions with one-particle states for the photoelectron, thus forming a so-called close-coupling space, and including in the space all those configurations obtained by adding one electron to an empty active orbital in any of the ionic configurations that form the MCHF ions. Full details on the configuration space and on the computation methodology are provided in our recent work on the numerical and analytical description of autoionizing polaritons~\cite{CarikerPRA2024}. As shown in~\cite{HarkemaPRL2021,CarikerPRA2024}, it is possible to identify, within the larger basis of close-coupling configuration functions, a small subset of essential states relevant to the calculation of optical observables, thus accelerating simulations by several orders of magnitude. Using the essential state basis, furthermore, we are also able to isolate and selectively exclude individual autoionizing states from the simulations, which is useful to identify their contribution to the spectrum. In the calculations, the resonance energies are shifted to match the known experimental values.

To compute the evolution of the wave function $|\Psi(t)\rangle$ of the atom under the influence of the external fields, we solve the time-dependent Schr\"odinger equation (TDSE), starting from the ground state of the atom, using a split-exponential second-order propagator, within the dipole approximation, in velocity gauge. For optically thin samples, the transient absorption spectrum (collinear signal) can be calculated as $\sigma(\omega) = \frac{4\pi}{\omega}\Im m\big[\tilde{p}(\omega)/\tilde{A}(\omega)\big]$, where $p(\omega)$ and $A(\omega)$ are the Fourier transform of the expectation value of the canonical momentum $p(t)=\langle \Psi(t)|\hat{p}_z|\Psi(t)\rangle$ and of the vector potential $A_z(t)$ of the XUV pulse, respectively. This formula is appropriate to estimate the main contributions to the transient absorption spectrum, which are due to the interference between the incoming XUV field, and the one generated by each atom, which are identical and hence add up coherently. 

The NCFWM signals, on the other hand, are due to the dipolar emission of the atoms alone, rather than to their interference with the incoming ionizing radiation (i.e., the signal is background free). Furthermore, to reproduce the non-trivial phase-matching conditions imposed by the angled probe, it is necessary to take into account the collective response of all the atoms in the interaction region, each of which experiences a slightly different delay between the collinear IR dressing pulse and the angled IR probe pulse. To do so, a sequence of two pump-single probe simulations are carried out for various values of the delay between the two probes, and the computed dipole response as a function of time delay is then used to evaluate the spatially resolved 4-wave mixing signal, as described in Appendix~\ref{app:NCFWMFormula}. The main approximations made when evaluating the collective response of the reaction region are that we assume that all the atoms in the interaction region experience the same laser intensity, and that the probe transition from the dark states to the $3s^{-1}6p$ state, which is the dominant contribution to the NCFWM signal, is linear in the intensity of both the collinear and non-collinear laser.

\section{Results~\label{sec:Results}}

In this section we compare the experimental measurements of the NCFWM signals illustrated in Fig.~\ref{fig:EnergyLevels} with the same quantities computed \emph{ab initio} as outlined in the previous section. In the proximity of the $3s^{-1}4p$ argon autoionizing state, both the experimental and theoretical ATAS spectra measured with IR peak intensities around $10$~GW/cm$^2$ or larger, such as those explored in this work, exhibit the hallmark of the Autler-Townes (AT) splitting due to the coupling between $3s^{-1}4p$ and the $3s^{-1}5s/3d$ dark states, which is indicative of Rabi cycling between these states~\cite{HarkemaPRL2021,CarikerPRA2024}. 
\begin{figure}[hbtp!]
\centering
\includegraphics[width=\columnwidth]{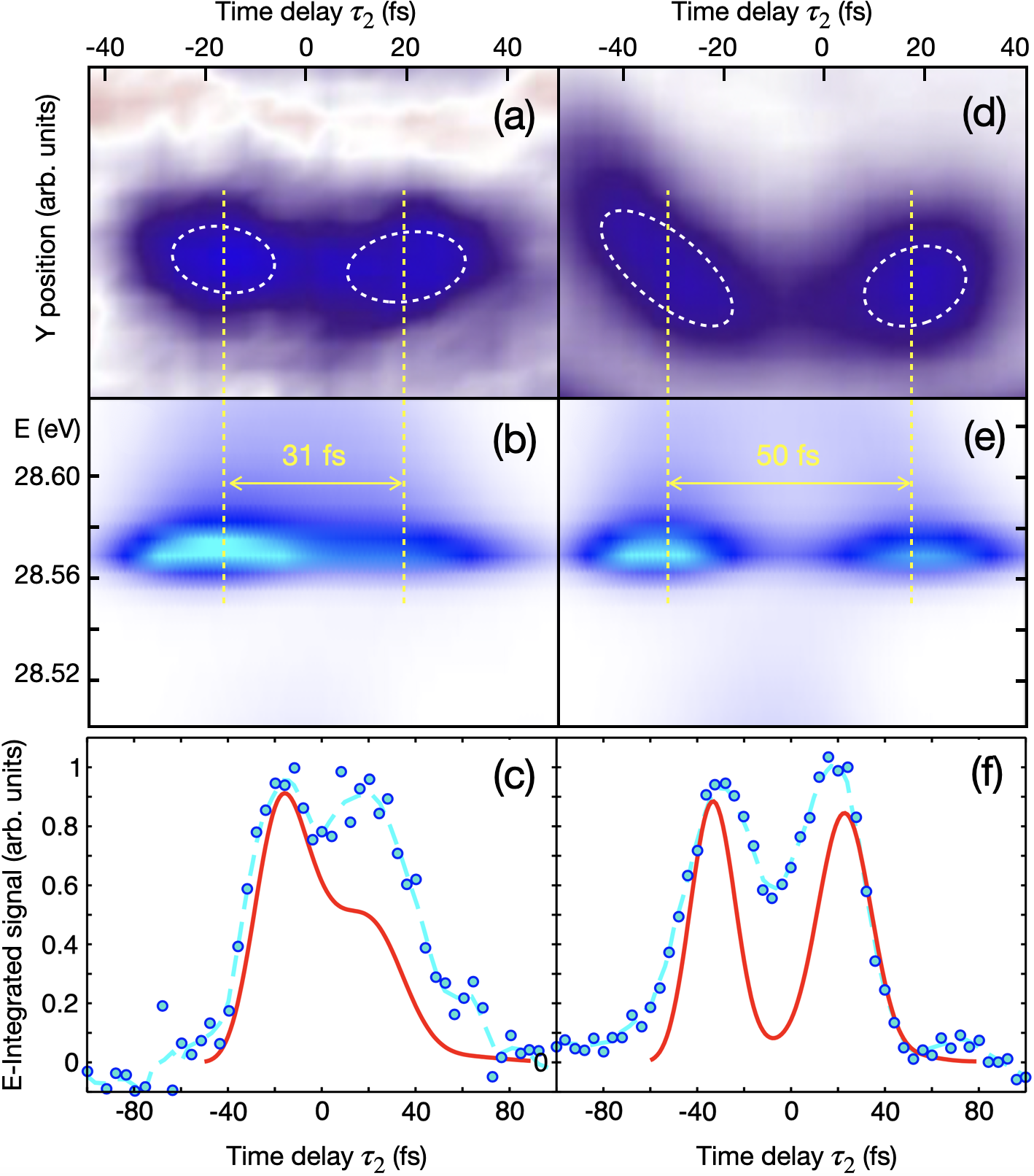}
\caption{\label{fig:ExpShift}(a,d) Experimental middle NCFWM signal in argon, as a function of time delay $\tau_2$ (horizontal axis) and emission angle (vertical axis), measured with the following nominal IR intensities: a) $I_{\textsc{ir}_1}=6.9~$GW/cm$^2$  and $I_{\textsc{ir}_2}=5.2$~GW/cm$^2$; d) $I_{\textsc{ir}_1}=37.6~$GW/cm$^2$  and $I_{\textsc{ir}_2}=19$~GW/cm$^2$. (b,e) Corresponding \emph{ab initio} signals, as a function of the emission frequency (vertical axis), evaluated at intensities that reproduce the Autler-Townes splitting of the ATAS signal: b) $I_{\textsc{ir}_1}=80~$GW/cm$^2$ and $I_{\textsc{ir}_2}=40$~GW/cm$^2$; e) $I_{\textsc{ir}_1}=160~$GW/cm$^2$ and $I_{\textsc{ir}_2}=80$~GW/cm$^2$. (c,f) Comparison between the experimental (circles, dashed cyan lines) and theoretical (red solid lines) energy-integrated signals; The data have been rescaled to fit between 0 and 1; The theoretical data have been horizontally shifted to better compare the separation between peaks.
}
\end{figure}

With the delay of the IR dressing pulse set to maximize the AT splitting, we measure and simulate the NCFWM signal associated to the net absorption of one photon from both the dressing and the angled IR probe pulses, as a function of the $\tau_2$ delay between the XUV and the angled probe pulse, for two different intensities of the probe lasers. This signal monitors the radiative transition from the $3s^{-1}5s/3d$ resonances to the $3s^{-1}6p$ bright state.
Figures \ref{fig:ExpShift}a shows the experimental signalwhen the angled probe pulse overlaps with the IR dressing pulse, with a nominal intensity for the collinear and non-collinear pulses of $6.9$ and $5.2$~GW/cm$^2$, respectively. Instead of following the convolution of the two IR intensity profiles, as one would expect from a non-sequential two-photon transition between to isolated states, the NCFWM signal is instead characterized by a pronounced minimum between two dominant peaks. At higher dressing-laser intensities, the separation between the two peaks increases, as shown in Fig.~\ref{fig:ExpShift}b, where the nominal intensity of the collinear and non-collinear pulses are $37.6$ and $19.0$~GW/cm$^2$, respectively. It is to be noted that the peak intensity of the lasers in the interaction region is difficult to estimate and is subject to considerable uncertainty.

To compute the NCFWM signal, we carry out several simulations in which the XUV and the collinear IR dressing pulses have a fixed relative delay whereas the intensity of both the collinear and non-collinear pulses are changed in fixed proportion. To calibrate the intensity of the pulses in the simulation, we match the Autler-Townes splitting of the $3s^{-1}4p$ signal, which is directly observable in the experiment~\cite{Yanez-PagansPRA2022}. Figure~\ref{fig:ExpShift}c shows the theoretical NCFWM signal computed at an intensity of 80~GW/cm$^2$ and 40~GW/cm$^2$ for the collinear and angled laser, whereas in Figure~\ref{fig:ExpShift}d we use 160~GW/cm$^2$ and 80~GW/cm$^2$, respectively. The XUV has a duration of 4~fs and intensity of 1~GW/cm$^2$, with energy tuned to the $3s4p$. All IR pulses have wavelength of 1350~nm and a 30~fs pulse duration. la{Figures~\ref{fig:ExpShift}e,f compare the experimental and theoretical energy-integrated signals from the panels above. 
The simulations reproduce the main features of the behavior observed in the experiment, namely, the formation of two peaks separated by a time-delay region where the NCFWM signal is suppressed, and the increase in the separation between the two peaks as the intensity of the two lasers is increased.} This overall good agreement confirms the validity of the perturbative approach used to evaluate the NCFWM signal, as described in App.~\ref{app:NCFWMFormula}, even when a few localized states are strongly coupled by a dressing field.

\begin{figure}[hbtp!]
\centering
\includegraphics[width=\columnwidth]{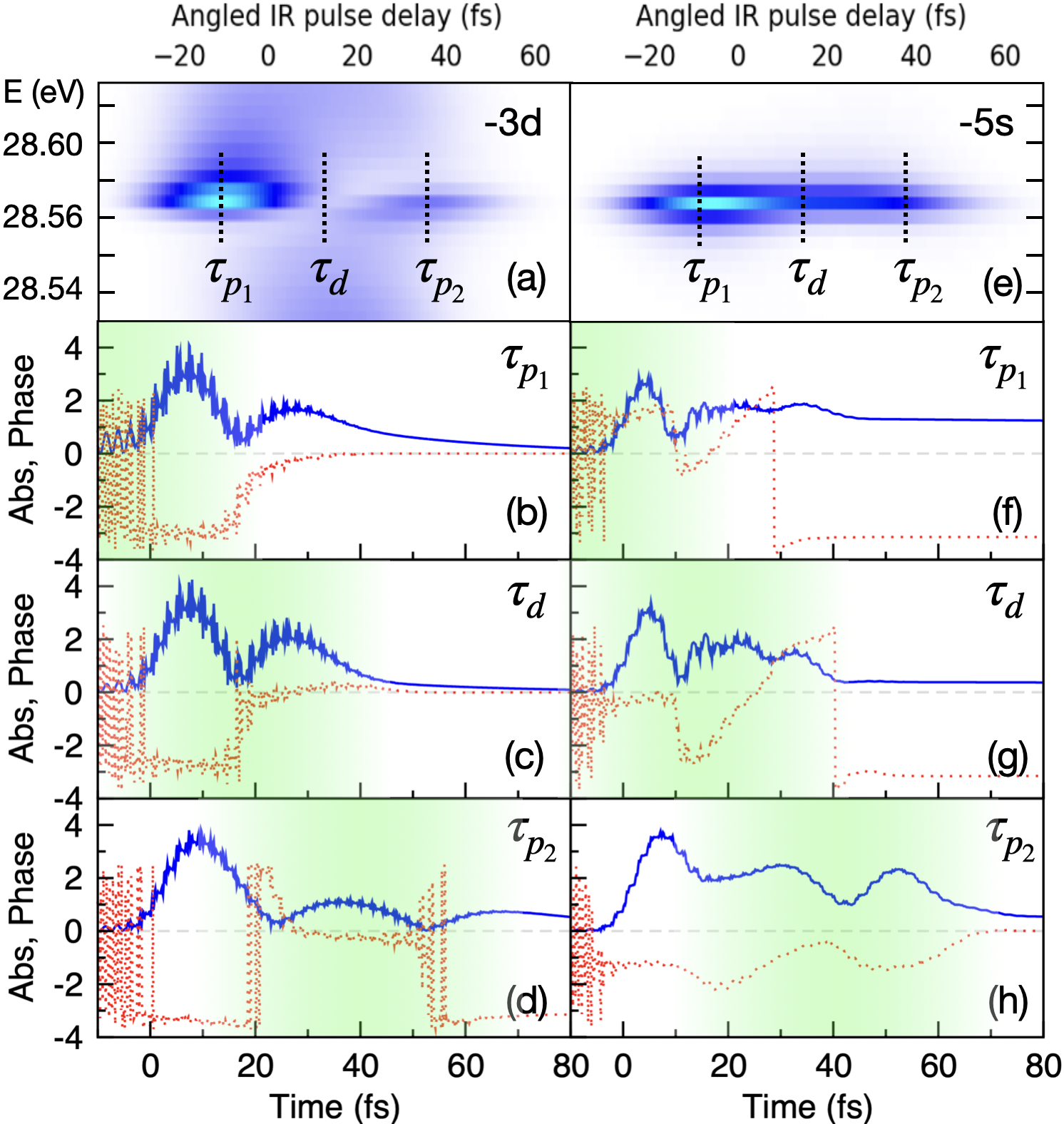}
\caption{\label{fig:Dark_AIS_Amplitudes} Left panels (a-d): simulations conducted excluding the $3s^{-1}3d$ state from the basis, with $I_{\textsc{IR}_1}=160$~GW/cm$^2$ and $I_{\textsc{IR}_2}=80$~GW/cm$^2$. Right panels (e-h): simulations conducted excluding the $3s^{-1}5s$ state from the basis. (a,e): NCFWM signal (compare with Fig.~\ref{fig:ExpShift}e). (b-d) Absolute value (blue solid lines) and phase (red dotted lines) of the $3s^{-1}5s$ dark autoionizing state, as a function of time, for three different time delays $\tau_2$ of the probe laser pulse (shaded regions), corresponding to the observation of (b) the first peak, at $\tau_{p_1}=-8.2$\,fs, (c) the trough, at $\tau_{t}=15$\,fs, and (d) the second peak, at $\tau_{p_2}=40$\,fs. (f-h) Same quantities for the $3s^{-1}3d$ state. The shaded region (green) illustrates the temporal range of the probe pulse $\textsc{IR}_2$. Both AIS undergo large Rabi oscillations. For negative delay, the probe overlap with the onset of the Rabi cycling, where both AIP's amplitudes have a same sign. For the middle time delay, on the other hand, the amplitude of both AIPs undergoes a sign change during the probe pulse, resulting in destructive interference, and hence in a reduced four-wave-mixing signal. At the larger delay, the amplitudes are once again of one sign throughout the probe pulse, resulting in a second peak in the signal.}
\end{figure}

To investigate the origin of the splitting, we repeated the simulations at $I_{\textsc{IR}_1}=80$~GW/cm$^2$ and $I_{\textsc{IR}_2}=40$~GW/cm$^2$ by omitting either of the two intermediate dark states from the configuration space. As the results in Fig.~\ref{fig:Dark_AIS_Amplitudes}a,e show, the bimodal structure of the NCFWM signal persists in both cases, even if the amplitude of the second peak is reduced.
The effect, therefore, is primarily due to the probe transition from either states and reflects the Rabi cycling of these two states with the $3s^{-1}4p$ resonance. Figure~\ref{fig:Dark_AIS_Amplitudes}b-d,f-h show the absolute value and the phase of the time-dependent coefficients of the $3s^{-1}5s$ (left column) and $3s^{-1}3d$ (right column) dark states, as a function of the simulation time, at three selected probe delays, $\tau_{p_1}$, $\tau_d$, and $\tau_{p_2}$, corresponding to the first peak, to the dip, and to the second peak in the NCFWM signal, respectively. To facilitate the interpretation of these plots, the coefficients are computed in the interaction representation of the field-free Hamiltonian and rigidly rotated in the complex plane such as their imaginary part vanishes at $\tau\to\infty$.
The support of the angled probe pulse is schematically indicated by a green shade. From these figures, it is clear that, for each dark state, the dip is caused by the destructive interference between the transition amplitudes generated at different times, as the state coefficient changes sign during the probe pulse. Conversely, the two peaks in the signal occur when the probe samples a dark state across a time intervale where its coefficient does not change sign. 
The dip is more pronounced for the $3s^{-1}5s$ state than for the $3s^{-1}3d$ state. This result is not surprising since the dynamics of the $3s^{-1}3d$ state is complicated by its strong coupling to the $3s^{-1}4f$ resonance, causing the minimum to be blurred. 
This effect can also be interpreted from a frequency perspective, as a consequence of the AT splitting of the $3s^{-1}5s/3d$ states into dark polaritons, generated by the coupling with $3s^{-1}4p$. As soon as half the AT splitting exceeds the probe bandwidth, the dressed dark state moves out of resonance from the upper $3s^{-1}6p$ state, resulting in the suppression of the NCFWM signal.


\section{Conclusions~\label{sec:Conclusions}}

In this work we have used NCFWM spectroscopy to investigate, both experimentally and theoretically, the non-perturbative radiative coupling between dark and bright autoionizing states of the laser-dressed argon atom. 
We find that the NCFWM signal corresponding to the $3s^{-1}3d/5s\to3s^{-1}6p$ transition is suppressed when the angled IR pulse probes the atom near the peak intensity of the dressing pulse, where the frequency of the Rabi beating between the $3s^{-1}3d/5s$ dark states and the $3s^{-1}4p$ bright state is largest, giving rise to an initial and a final peak in the signal that correspond to the onset and to the end of the Rabi oscillations. This interpretation is consistent with the distance between the two peaks increasing with the intensity of the dressing laser, and it is supported by \emph{ab initio} simulations.
In perspective, increased resolution and statistics should allow us to leverage NCFWM transitions to the non-resonant continuum to image in full the Autler-Townes splitting of dark autoionizing states, which is too elusive in collinear transient absorption spectra~\cite{ArgentiPRA2015}.

\section{Acknowledgements}
We gratefully acknowledge the support received for this work. AS, NH, and SP acknowledge the support from the U.S. Department of Energy, Office of Science, Basic Energy Sciences, under Award No.~DE-SC0018251. LA and CC thank the National Science Foundation's Theoretical AMO Physics program for their support through grants No.~1607588 and No.~1912507. The authors are grateful for the computational resources provided by the UCF ARCC. 

\appendix

\section{\label{app:NCFWMFormula}\emph{ab initio} Evaluation of NCFWM Response}
To relate the single-atom response to the collective response of the interaction region, we note that each atom in the region experiences a sequence of one XUV pump and two IR probe pulses. Let $\hat{x}$ be the propagation direction of the XUV and IR$_1$ pulses and $\hat{y}$ the vector normal to $\hat{x}$ in the plane defined by the propagation direction of IR$_1$ and IR$_2$, which form an angle $\theta$ with each other. The origin of our system of reference is chosen at the intersection of the two IR beam axes. The remaining direction, $\hat{z}$, is perpendicular to the propagation plane and parallel to the polarization of all pulses, and hence it is parallel also to the induced atomic dipoles throughout the interaction region.
Starting from the Li{\'e}nard-Wiechert potentials~\cite{Jackson1998}, it can be shown that a time-dependent dipole $\hat{z}\,{p}(\vec{r}',t')$ gives rise to an electric field $\vec{\mathcal{E}}(\vec{r},t)$ whose spectrum, at a distant point $\vec{r}=r\hat{k}$, in proximity of the $\hat{x}$ axis $(1-\hat{k}\cdot\hat{x}\ll 1)$ and to leading order in $1/r$, is proportional to
\begin{equation}
\vec{\mathcal{E}}(\vec{r},\omega)\propto \hat{z}\,p(\vec{r}',\omega)\, \frac{\omega}{r}\,\exp\left[i\frac{\omega}{c}(x'-\hat{k}\cdot\vec{r}')\right],
\end{equation}
where $f(\omega)=\frac{1}{2\pi}\int dt f(t) e^{i\omega t}$, we have assumed that the charged particles giving rise to the dipole move much more slowly than the speed of light, and that $r\omega \gg c$.
The contribution to $\mathcal{E}_z(\vec{r},\omega)$ of the dipoles in an infinitesimal element of volume $d^3r'$ in the interaction region, with numerical density $\rho(\vec{r}')$ is
\begin{equation}\begin{split}
d\mathcal{E}_z(\vec{r},\omega) &\propto \rho(\vec{r}') d^3r' e^{\frac{i\omega}{c} (x'-\hat{k}\cdotp\vec{r}')}\,\,p_z(\vec{r}',\omega;\tau_1,\tau_2),
\end{split}\end{equation}
where $\tau_1$ and $\tau_2$ are the delays with which the peak of IR$_1$ and IR$_2$ cross the origin, relative to the XUV, and $p_z(\vec{r}',\omega;\tau_1,\tau_2)$ is the dipole moment spectrum of an atom at $\vec{r}'$ in an experiment with nominal delays $\tau_1$ and $\tau_2$.
While the apparent delay $\tau_1'=t_{IR_1}(\vec{r}')-t_{XUV}(\vec{r}')=\tau_1$ between the IR dressing pulse (IR$_1$) and the collinear XUV pulse is the same for all atoms, the apparent delay $\tau_2'=t_{IR_2}(\vec{r}')-t_{XUV}(\vec{r}')$ of the angled probe pulse (IR$_2$) depends on the position of the atom within the region.  
The coordinate of a point $(x,y,z)$ along the propagation direction of IR$_2$ is $x_2=x\cos\theta+y\sin\theta$. An atom centered at $(x,y,z)$ therefore, experiences IR$_2$ with the following apparent delay with respect to the XUV pulse 
\begin{equation}
\begin{split}
\tau_2(x,y)&=t_{\mathrm{IR}_2}(x,y)- t_{XUV}(x) =\\
&=\tau_{2}(0,0)+(\cos\theta-1)\frac{x}{c}+\frac{y}{c}\sin\theta\simeq\\
&\simeq \tau_{2}(0,0)+\frac{y}{c}\theta,
\end{split}
\end{equation} 
where we have assumed $\theta\ll 1$ and retained only the term linear in $\theta$. Thus, we can rewrite the contribution of an atom at a position $\vec{r}'$ to the emitted electric field in terms of the response of an atom at the origin, in a pump-probe simulation with effective time delays $\tau_1$ and $\tau_2'=\tau_2(\vec{r}')$, where the peak amplitudes of the pump and probe pulses, $A_{\textsc{xuv}}$, $A_{\textsc{IR}_1}$, and $A_{\textsc{IR}_2}$,  are those measured at $\vec{r}'$,
\begin{equation}\begin{split}
d\mathcal{E}_z(\vec{r},\omega) &\propto \rho(\vec{r}')\, d^3r'\, e^{\frac{i\omega}{c} (x'-\hat{k}\cdotp\vec{r}')}\,\times\\
&\times\,p_z\left[\omega;\tau_1,\tau_2+y\theta/c,\big\{\vec{A}_j(\vec{r'})\big\}\right].
\end{split}\end{equation}
The total spectrum of the electric field is the integral of $d\mathcal{E}_z(\vec{r},\omega)$ over the whole interaction region
\begin{equation}\begin{split}
\mathcal{E}_z(\vec{r},\omega) &\propto \int \rho(\vec{r}')\, d^3r'\, e^{\frac{i\omega}{c} (x'-\hat{k}\cdotp\vec{r}')}\,\times\\
&\times\,p_z\left[\omega;\tau_1,\tau_2+y\theta/c,\big\{\vec{A}_j(\vec{r'})\big\}\right].
\end{split}\end{equation}
To carry out this integral, we assume that the beams have a Gaussian intensity profile in the plane perpendicular to their direction of propagation, whereas the intensity is constant along the direction of propagation,
\begin{equation}
\begin{split}
A_{\textsc{xuv}}(\vec{r}) = A_{\textsc{xuv}}(\vec{0})\exp\left(-\frac{y^2+z^2}{w_{XUV}^2}\right), \\
A_{\textsc{ir}_i}(\vec{r}) = A_{\textsc{ir}_i}(\vec{0})\exp\left(-\frac{y^2+z^2}{w_{\textsc{ir}_i}^2}\right).
\end{split}
\end{equation}
For the gas particles, we similarly assume density constant in the $\hat{y}$ and $\hat{z}$ directions with a Gaussian profile along $\hat{x}$, 
\begin{equation}
\rho(\vec{r}') \propto \exp\left(-\frac{x^2}{w^2_g}\right),
\end{equation}
which approximates the experimental conditions. 

While the response of the atom is strictly linear with respect to the XUV field, which is extremely weak, and possibly with respect to the angled probe IR pulse as well, it is certainly not linear with respect to the IR dressing pulse, due to the strong coupling the latter induces between autoionizing states. A proper volume averaging, therefore, would require to evaluate the single-atom response not only for many different apparent delays $\tau_2'$, but also for several intensities of the dressing laser. In the present work, however, we are interested in the volume averaging for the main purpose of determining the non-collinear emission, rather than to mimic the circumstance of different atoms experiencing different peak intensities. In fact, the latter conditions affects already the ATAS spectrum, which we have already ascertained being satisfactorily reproduced by simulating a single-atom response to a single nominal intensity. Here we will similarly assume a same peak intensity for all the simulations. However, we will also assume that the NCFWM signal of the laser-dressed atom, corresponding to the net absorption of one IR$_1$ and one IR$_2$ photon, is linear in the amplitude of these two fields. This is not an unreasonable assumption, as the one-photon absorption from a dark state to the continuum, which is responsible for the NCFWM signal, is approximately linear. 
If we further neglect the $z$-dependence, which is immaterial here, we obtain an expression for the NCFWM signal along the phase-matching direction, $\vec{k}=\vec{k}_{\textsc{xuv}}+\vec{k}_{\textsc{ir}_1}+\vec{k}_{\textsc{ir}_2}$ as a numerical integral of the dipole response spectrum over a range of simulated time delays,
\begin{equation}\nonumber
\begin{split}
 \mathcal{E}_z^{\textsc{ncfwm}}&(\omega;\tau_1,\tau_2) \propto
  A_{\textsc{ir}_1}A_{\textsc{ir}_2} \times\\
  &\times \int du' \exp\left[
-A\,u'^2 -B\,u'
  \right] p_z\left(\omega;\tau_1,\tau_2+\frac{\theta}{c}u'\right),
  \end{split}
 \end{equation}
where $x'=v'-\frac{\theta}{2}u', y'=u'+\frac{\theta}{2}v'$, and we introduced the following constants 
\begin{eqnarray}
&&A=\frac{1}{w_{XUV}^2}+\frac{1}{w_{IR1}^2}+\frac{1+\theta^2}{w_{IR2}^2}-\frac{D^2}{C},\nonumber\\
&&B=\frac{i\theta\omega_{IR}}{c}+\frac{i\beta D}{C},\quad\beta=\frac{\theta^2\omega_{IR}}{2c},\quad C=\frac{1}{w_g^2}-\frac{\theta^2}{w_{IR2}^2},\nonumber\\
&&D=-\frac{\theta}{2}\left(
 \frac{1}{w_{XUV}^2}+\frac{1}{w_{IR1}^2}-\frac{1}{w_{IR2}^2}-\frac{1}{w_g^2} \right).\nonumber
\end{eqnarray}
The quantity to be compared with the experiment, reproduced in the next section, is the square module of $\mathcal{E}_z^{\textsc{ncfwm}}$.

\bibliographystyle{apsrev4-2}
\bibliography{library_LA} 

\end{document}